\documentclass[nofootinbib, superscriptaddress, amsmath,amssymb, aps, twocolumn ]{revtex4-1}

\usepackage{graphicx}
\usepackage{ulem}
\usepackage{dcolumn}
\usepackage{bm}
\usepackage{wasysym}
\usepackage{hyperref}
\usepackage[mathlines]{lineno}
\usepackage{xcolor}
\usepackage{dsfont}

\usepackage{gensymb}
\usepackage{enumitem}
\usepackage{multirow}
\usepackage{booktabs}
\usepackage{array}
\newcolumntype{P}[1]{>{\centering\arraybackslash}p{#1}}
\usepackage{amssymb}
\usepackage{adjustbox}
\usepackage{orcidlink}
\usepackage{float}
\usepackage{graphicx}
\usepackage{dcolumn}
\usepackage[caption=false]{subfig}
\usepackage{tikz,fp}
\usetikzlibrary{external,fixedpointarithmetic,decorations.pathmorphing,positioning,calc,fadings,decorations.markings,decorations.pathreplacing,patterns,arrows}

\tikzset{
  label distance=-3pt,
  >=stealth,
  inner sep=1.5pt,
  boson/.style={
    decoration={snake, segment length=2mm, amplitude=0.5mm},
    decorate,
  },
  scalar/.style={
    dashed
  }
}

\newcommand*{\defeq}{\mathrel{\vcenter{\baselineskip0.5ex \lineskiplimit0pt
                     \hbox{\scriptsize.}\hbox{\scriptsize.}}}
                     =}

\newcommand{\iu}{\mathrm{i}\mkern1mu}
\newcommand{\du}{\mathrm{d}}

\begin{document}

\title{Gravitational scattering of solitonic boson stars: Analytics vs Numerics}

\author{Thibault Damour} 
\email{damour@ihes.fr}
\affiliation{Institut des Hautes Etudes Scientifiques,
91440 Bures-sur-Yvette, France}

\author{Tamanna Jain} 
\email{tj317@cam.ac.uk}
\affiliation{LPENS, D{\'e}partement de physique, Ecole normale sup{\'e}rieure,
Universit{\'e} PSL,
Sorbonne Universit{\'e}, Universit{\'e} Paris Cit{\'e}, CNRS, 75005 Paris}
\affiliation{Institut des Hautes Etudes Scientifiques, 
91440 Bures-sur-Yvette, France}
\affiliation{DAMTP, Centre for Mathematical Sciences,
University of Cambridge, Wilberforce Road, Cambridge CB3 0WA, UK}

\author{Ulrich Sperhake.
\orcidlink{0000-0002-3134-7088}}
\email{U.Sperhake@damtp.cam.ac.uk}
\affiliation{DAMTP, Centre for Mathematical Sciences,
University of Cambridge, Wilberforce Road, Cambridge CB3 0WA, UK}
\affiliation{Department of Physics and Astronomy, Johns Hopkins University,
3400 North Charles Street, Baltimore, Maryland 21218, USA}
\affiliation{{TAPIR 350-17, Caltech, 1200 E. California Boulevard,
Pasadena, California 91125, USA}}

\date{\today}


\begin{abstract}
We study the scattering of boson-star binaries, taking into account
three effects: point-mass gravitational, tidal, and short-range
scalar-field interactions. We compare analytic results to the
scattering angle extracted from four sequences of numerical-relativity
simulations at fixed energy and varying impact parameter. The
very good agreement exhibits the attractive (repulsive)
effect of in-phase (out-of-phase) binaries, wich dominates at small impact parameters. We thus obtain the first effective-one-body potential, central for the construction of analytic gravitational-wave templates. 

\end{abstract}

\maketitle

{\textit{\textbf{Introduction}}---}
The first direct detection of a gravitational wave (GW) signal,
GW150914, from a black-hole (BH) binary~\cite{LIGOScientific:2016aoc},
followed by $\mathcal{O}(100)$ further detections to date
\cite{KAGRA:2021vkt,LIGOScientific:2025hdt,LIGOScientific:2025yae,LIGOScientific:2025slb}, has established gravitational-wave astronomy
as a new pathway for exploring the physics of our universe.  New
tests of general relativity (GR) have added more evidence supporting
the exceptional accuracy and range of validity of Einstein's theory
\cite{LIGOScientific:2020tif,LIGOScientific:2021sio,LIGOScientific:2025rid}.  The multi-messenger observation of
the neutron-star (NS) binary system GW170817 \cite{LIGOScientific:2017vwq}
cemented a strong relation to short gamma ray bursts and efficient
production of heavy elements.

While all events observed so far are compatible with compact binaries
composed of NSs and BHs, the mass values inferred
from some events raise intriguing questions.  The analyses of
GW190521 and GW231123 suggest constituent black holes (BHs) with
masses $\gtrsim 85\,M_{\odot}$ inside or above the pair-instability
induced mass gap \cite{KAGRA:2021vkt,LIGOScientific:2025rsn}, while
the secondary of GW170814 with a mass $2.50-2.67\,M_{\odot}$ would
either be the heaviest NS or the lightest BH observed.
These observations have naturally triggered investigations into the
possibility of alternative, more exotic origins of these events as
for example cosmic strings \cite{Aurrekoetxea:2023vtp}, Proca stars
\cite{CalderonBustillo:2020fyi} or primordial black holes
\cite{Clesse:2020ghq}. Furthermore injections of inspiral-merger-ringdown
waveforms from boson stars (BSs) into simulated LIGO noise have
revealed significant degeneracy between signals from BS and BH
binaries \cite{Evstafyeva:2024qvp}. Such exotic GW sources
have attracted a lot of attention in diverse contexts as dark
matter searches \cite{Liebling:2012fv,Kim:2024wku}, structure
formation \cite{Vachaspati:1991tt,Inman:2019wvr}, BH mimickers
\cite{Cunha:2022gde,Siemonsen:2024snb,Marks:2025jpt,Evstafyeva:2025mvx} or
the validity of Thorne's hoop conjecture \cite{Choptuik:2009ww}.

Clearly, the capacity of GW observations to distinguish between
BHs, NSs and exotic compact objects will be crucial
for identifying the nature of as yet unclassified sources, to
understand their astrophysical origin, and to search for new physics
beyond the standard model of particles. Systematic investigations
are naturally drawn to two main characteristics of these diverse
sources, (i) tidal interactions (e.g.~\cite{Gonzalez:2025xba}) and
(ii) manifestations in either GW emission or the electromagnetic
spectrum arising from the specific matter they are composed of
(cf.~\cite{Palenzuela:2017kcg} or the electromagnetic counter parts
of GW170817).

In this work, we consider such effects (including the additional
short-range scalar-field interaction) for the case of BS binary
systems. A variety of stellar configurations composed of bosonic
matter has been studied in the literature, as for example Proca
stars \cite{Brito:2015pxa} or $\ell$ BSs \cite{Alcubierre:2018ahf}.
Here we focus on the class of BSs composed of a single complex
scalar field first identified by Kaup \cite{Kaup:1968zz}; see also
\cite{Ruffini:1969qy}, and Refs.~\cite{Liebling:2012fv,Visinelli:2021uve}
for reviews.  We treat BSs as electromagnetically dark, so that
direct effects of their scalar matter will manifest themselves
exclusively through overlap in the regime where the two BSs are in
close vicinity.  Tidal effects, on the other hand, will be present
at any stage of the binary evolution, albeit with decreasing magnitude
at larger distance.

In this work, we compute for BS binaries the effective 
potential central for the construction of analytic GW templates
using the effective-one-body (EOB)
formalism~\cite{Buonanno:1998gg}, which in itself uses results of
various perturbative approaches: post-Newtonian (PN), post-Minkowskian
(PM), Self-Force, Effective Field Theory (EFT) and Tutti Frutti.
We furthermore employ numerical results which are necessary to determine
the accuracy and validity of these approximants used to construct
semi-analytic banks of waveform templates essential for detection
and parameter estimation of GW events.

Whilst numerical relativity (NR) simulations of binary BSs have
been performed for quasi-circular orbits, in this work we present
the first simulations of binary BS scattering along with analytically
deriving for the first time the effect of scalar interactions for
binary BSs. We then compare the scattering angle $\chi$
determined from NR to the analytical estimates from an EOB-potential model
including: point-particle results (BH), tidal effects
and scalar field interactions. We find that the scalar effect
dominates for small impact parameters while the tidal effects are
sub-dominant; nonetheless, the inclusion of tidal effects order by
order improves the agreement with NR data.

We use a ``mostly plus'' signature for the spacetime metric and employ
units where the speed of light $c=1$.

{\textit{\textbf{Theory}}---}
The action describing the dynamics of a complex scalar field
$\varphi$  minimally coupled to gravity is,
\begin{equation}
  \!\!S = \int \frac{\sqrt{-g}}{2}\left\{ \frac{R}{8\pi G}
        -\left[ g^{\mu\nu}\nabla_{\mu}{\bar{\varphi}}
        \nabla_{\nu}\varphi
        +V(|\varphi|)\right]\right\}\du^4x,
        \label{eq:action}
\end{equation}
where $R$ is the Ricci scalar, $g$ the determinant of the spacetime
metric $g_{\mu\nu}$, an overbar denotes the complex conjugate and
$V(|\varphi|)$ is the scalar-field potential. In this work, we
consider solitonic potentials characterized by the scalar mass $\mu$
(with dimension of inverse length) and a parameter $\sigma_0$,
\begin{equation}
  V(|\varphi|) = \mu^2 |\varphi|^2\left(1
  - 2 |\varphi|^2/\sigma_0^2\right)^2\,.
\end{equation}
Henceforth we use $\sqrt{G} \sigma_0=0.2$, which accommodates BS
solutions with high compactness between that of NSs and
BHs.  The coupled Einstein-Klein-Gordon equations governing
the BS solution are obtained by varying the action \eqref{eq:action}.
Spherically symmetric solutions are obtained with the harmonic
ansatz~\cite{Carloni:2019cyo}, $\varphi (t,r) = A(r) e^{\iu (\epsilon
\omega t+ \Phi)}$, where $A(r)$ is the (real) scalar field
amplitude profile, $\omega$ is the frequency, and $\Phi$ is
an initial phase. 
We also introduce a parameter $\epsilon =
\pm1$ determining the sense of rotation of the scalar field in the
complex plane. By convention, $\epsilon=1$ defines a BS while
$\epsilon = -1$ is an {\it antiboson star} ($\overline{\text{BS}}$).
Single BS and $\overline{\text{BS}}$ spacetimes are identical except
for opposite Noether charge densities.

For a given value of the central scalar-field amplitude $A_{\rm
ctr} = A(0)$, the Einstein-Klein-Gordon equations along with
boundary conditions for asymptotic flatness {(typically)} admit a unique
ground-state BS (or $\overline{\text{BS}}$) solution. In this
work, we consider $\sqrt{G}A_{\rm ctr}=0.17$ which corresponds
to a BS containing 99\% of its total mass $m$ within a radius
$r_{99} = 5.58\,Gm$ ($=3.98\mu^{-1}$).

\begin{table}[t]
\caption
{Impact parameter $b$, total angular momentum $J_{\rm NR}$,
and the
scattering angles (together with their error bars) for the four BS
binary sequences as extracted from the numerical simulations. For
all configurations, the total energy is $E/M = 1.02394$.
}
\begin{center}
\footnotesize{
\begin{tabular}{  c   c  c  c   c  c }
\hline
$\frac{b}{GM}$ & $\frac{{J_{\rm NR}}}{GM^2}$ & $\chi^{\rm BSBS}_{\rm NR}\,[{}^{\circ}]$ & $\chi^{\rm BS\overline{BS}}_{\rm NR}\,[{}^{\circ}]$ & $\chi^{\rm BSBS^{\frac{\pi}{2}}}_{\rm NR}\,[{}^{\circ}]$ & $\chi^{\rm BSBS^{\pi}}_{\rm NR}\,[{}^{\circ}]$ \\
\hline
\hline
9.9 & 1.15315   & 258.79(95) & 216.41(1.41) & 214.18(1.40) & 198.41(1.26) \\
10.5 & 1.22360  & 170.51(89) & 165.42(81) & 165.28(81) & 161.23(75) \\
11.0 & 1.28186  & 143.91(62) & 142.32(60) & 142.28(60) & 140.83(58)\\
12.0 & 1.39839 & 114.43(64) & 114.20(63) & 114.33(53) & 113.96(65)\\
13.0 & 1.51492  & 96.89(84) & 96.89(84) & 96.75(60) & 96.81(82)\\
14.1 & 1.64982  & 83.39(1.14)& 83.38(1.14) & 83.32(1.10) & 83.32(1.11) \\
15.0 & 1.74798  & 75.51(1.23) & 75.73(1.37) & 75.62(1.30) & 75.62(1.30) \\
16.0 & 1.86450 & 68.16(1.27)& 68.44(1.44) & 67.97(1.14) & 68.50(1.47) \\
\hline
\end{tabular}

\vspace{-0.5cm}
}
\label{tab:initData}
\end{center}
\end{table}

{\textit{\textbf{Numerical Relativity (NR) Simulations}}---}
Our BS binary simulations have been performed using the \textsc{lean}
code \cite{Sperhake:2006cy} based on the {\sc cactus computational
toolkit} \cite{Allen:1999} using 4th-order finite differencing of
the CCZ4 \cite{Alic:2011gg} version of the Einstein-Klein-Gordon
equations. The code employs the moving puncture gauge and utilizes
mesh refinement via {\sc carpet} \cite{Schnetter:2003rb}. Our
computational domain extends to $448\,GM$, where $M\defeq m_1+m_2=2m$
is the total BS rest mass, using 8 nested refinement levels with
standard resolution $\Delta x=0.02\,GM$ on the innermost level. We
employ outgoing Sommerfeld boundary conditions.

We study the scattering of two equal-mass ($\nu\defeq
m_1m_2/M^2=1/4$), nonspinning  BSs with scalar frequency
$\omega=\omega_1=\omega_2$. We investigate four distinct binary
systems {composed of a primary BS with} fixed parameters ($\epsilon
= 1$, $ \Phi = 0$), and a secondary with parameters varying
as follows: (i) ($\epsilon = 1$, $ \Phi = 0$), denoted BSBS;
(ii) ($\epsilon = 1$, $ \Phi = \pi/2$), denoted BSBS$^{\frac{\pi}{2}}$;
(iii) ($\epsilon = 1$, $ \Phi = \pi$), denoted BSBS$^\pi$;
and (iv) ($\epsilon = -1$, $ \Phi = 0$), denoted
BS$\overline{\text{BS}}$.  The BSs start on the $x$-axis with
initial positions $\pm X$ and initial velocity parameters $\vec{v}=(\mp
v_x,\pm v_y,v_z)$, with corresponding kinematic momenta $\vec{p}
\defeq m\vec{v}/\sqrt{1-\vec{v}^2}$. Unless specified otherwise,
we set $|\vec{v}| = 0.2291$ and $X=50.49\,GM$.  Defining the impact
parameter $b\defeq 2Xv_y/|\vec{v}| = 2Xp_y/|\vec{p}|$, the velocity
vector of the two BSs is given by $(v_x, v_y, v_z ) = \pm |\vec{v}|
(-\sqrt{1 - (b/(2X))^2},b/(2X), 0)$.

Initial data is constructed using the improved superposition of two
boosted single-BS spacetimes as detailed in Ref.~\cite{Helfer:2021brt}.
Residual constraint violations inherent to these data are further
reduced by employing the CCZ4 formulation with constraint damping
parameter $\kappa_1=0.1/M$ following Ref.~\cite{Alic:2011gg}.  We
compute the binaries' initial angular momentum from Eq.~(\ref{deltaJ})
below as $J=(1+\epsilon_J)J_{\rm NR}$, where $J_{\rm NR}$ is the
Arnowitt-Deset-Misner (ADM) like integral (7.63) of
Ref.~\cite{Gourgoulhon:2007ue} extrapolated to infinite extraction
radius.  The initial energy $E$ is estimated as the sum of the
relativistic (rest $+$ kinetic) energy and the Newtonian binding
energy including a parameter $c_V$ for relativistic corrections,
\begin{equation} \label{EN}
  E = 2\frac{m}{\sqrt{1-v^2}}-c_V\frac{G m^2}{2 X}\,;
\end{equation}
{see text below, and the Appendix \ref{append1}, for estimates of $\epsilon_J$ and $c_V$, and   their justification.}

We adopt a method similar to that of
Refs.~\cite{Damour:2014afa,Hopper:2022rwo} to determine the scattering
angle. After converting the BS trajectories into standard polar
coordinates $(r,\theta)$ and fitting the dependence of the angles
on $1/r$ by polynomials $\theta_p\defeq \sum_{n=0}^p \theta_{n}^{p}/r^n$,
the angles (for each resolution) are obtained from the average
$\bar{\theta}= (\theta_0^2+\theta_0^3+\theta_0^4+\theta_0^5)/4$.
As a conservative estimate of the error due to polynomial fitting,
we use half the range of $\theta_0^p$ with $p=1,\ldots,6$, yielding
an uncertainty of $1.5{\degree}$ or less. A convergence analysis using
additional resolutions $\Delta x = 0.022\,GM$ and $0.017\,GM$ and
employing a fifth-order polynomial fit of the trajectory for the
binary configuration $b=9.9\, GM$ yields second-order convergence
with a discretization error $\sim 0.5\degree$.  Adding these two
values in quadrature gives us a combined error budget $\Delta \chi_{\rm NR} \sim 1.6{\degree}$ or less. The results so obtained are displayed in
Table~\ref{tab:initData}.

{\textit{\textbf{Analytical Relativity (AR) computation of $\chi$}}---}

Recently, there has been a significant effort on comparing analytic
results to NR simulations of BH scattering
\cite{Hopper:2022rwo,Damour:2022ybd,Rettegno:2023ghr,Swain:2024ngs,Buonanno:2024vkx,Long:2025nmj}.
Following Ref.~\cite{Damour:2022ybd}, we analytically describe the
scattering of BSs by transcribing the PM-expanded scattering angle
$ \chi(\gamma,j) = \sum_n  2 \chi_n(\gamma)/j^n$ (where
$j\defeq J/(Gm_1m_2)$)  into a corresponding energy-dependent EOB
radial potential $w(\gamma,\bar{r})= \sum_n  w_n(\gamma)/\bar{r}^n$ (where $\bar{r} \defeq \bar{R}/(GM)$ is a
dimensionless isotropic EOB radial coordinate). Here, and in the
following, $\gamma$ denotes the EOB effective energy of the system,
which is related to the total center-of-mass energy
$E \defeq M h$ via $h =\sqrt{1+ 2 \nu (\gamma-1)}$ or, equivalently
(for the equal-mass case $\nu=1/4$), $\gamma= 2 (E/M)^2-1$. [For scattering states, $\gamma$ happens to be equal to the
Lorentz factor between the two incoming world lines.]  The dynamics
of BS binaries involve two additional effects compared to those of
BHs: tidal effects and scalar-field interactions, so that the
PM-expansion of $\chi$ is naturally decomposed as
\begin{align}
  \label{eq:chitotal}
  \chi (\gamma, j) = \chi^{\rm BH}(\gamma,j)+\chi^{\rm tid}(\gamma,j)
  +\chi^{\rm scalar}(\gamma,j)~,
\end{align}
leading to a corresponding decomposition of the radial potential,
\begin{align}
  \label{eq:wdecompose}
  w (\gamma,\bar{r}) = w^{\rm BH} (\gamma,\bar{r}) + w^{\rm tid}
  (\gamma,\bar{r}) + w^{\rm scalar} (\gamma,\bar{r})~.
\end{align}
The PM expansion of $w$ is uniquely determined by the PM expansion of $\chi$ via the EOB-derived map \cite{Damour:2017zjx}
\begin{align}
  \label{eq:chirelation}
  \pi+\chi(\gamma,j) = 2 j \int_0^{\bar{u}_{\rm max}(\gamma, j)}
  \!\!\!\!
  \frac{d\bar{u}}{\sqrt{p_{\infty}^2+w(\bar{u},\gamma)-j^2 \bar{u}^2}}~,
\end{align}
where $\bar{u}=1/\bar{r}$\,, $\bar{u}_{\rm max} \equiv 1/\bar{r}_{\rm
min}$ and $p_{\infty}^2=\gamma^2-1$.

Based on the 4PM knowledge of $\chi^{\rm BH}$  \cite{Kalin:2020lmz},
$w^{\rm BH}$ is fully known analytically up to 4PM order
\cite{Damour:2022ybd}, and has been completed by a numerically
fitted 5PM contribution, $w_5(\gamma)^{\rm BH, NR}$ \cite{Rettegno:2023ghr}
-- recall that $w^{\rm BH}$ is an effective potential that includes
radiation-reaction effects \footnote{However, the overall recoil has to be included separately~\cite{Bini:2021gat}.}.  The tidal potential $w_n^{\rm tid}$
starts at 6PM, and is computed from the scattering angles derived
in Refs.~\cite{Bini:2020flp,Cheung:2020sdj, Kalin:2020lmz} using
the following consequences of Eq.~\eqref{eq:chirelation},
\begin{align} \label{chitidal}
  &\!\!\chi_6^{\rm tid} = \frac{15\pi}{32}p_{\infty}^4 w_6^{\rm tid}\,,
  ~~\chi_7^{\rm tid} = 4p_{\infty}^3 w_1 w_6^{\rm tid}+\frac{8}{5}
  p_{\infty}^5 w_7^{\rm tid}\,.
\end{align}

The last contribution to the potential, $ w^{\rm scalar}$, which
describes the short-range scalar interaction between two BSs, has
not been previously derived, nor its corresponding scattering-angle
contribution part, $ \chi^{\rm scalar}$.
We have computed the leading-order (LO) PM analytical value of
$w^{\rm scalar}$ describing the conservative part of the scattering for BS binary, namely
\begin{align} 
   w_{\rm LO}^{\rm scalar}= \frac{8\pi c_1 c_2}
   {G m_1 m_2} \frac{e^{ - \bar{m} \bar{r}}}{\bar{r}} \cos(\Phi_{21} )~.
   \label{eq:wscalar}
\end{align}
Here, $c_1$ and $c_2$ are real scalar charges (defined in the
next section),  $\bar{m} \equiv GM h \widetilde{\mu}$ [with $\widetilde{\mu}^2 \equiv \mu^2- (2
\omega_1\omega_2\gamma-\omega_1^2-\omega_2^2)/(\gamma^2-1)$] is an inverse range, and $ \Phi_{21} \defeq  \Phi_2- \Phi_1$ is the dephasing between the two
interacting BSs' scalar fields.  This corresponds to an attractive
potential for BSBS ($\Phi_{21} =0$), a repulsive one for BSBS$^\pi$
($\Phi_{21} =\pi$), and a vanishing one for BSBS$^{\frac{\pi}{2}}$
($\Phi_{21} =\pi/2$). For the BS${\overline {\rm BS}}$ case we
obtain a fast-oscillating potential which averages to zero.  These
results confirm in a quantitative way the qualitative considerations
of Appendix B of Ref.~\cite{Palenzuela:2006wp}.

We sketch the derivation of our results as follows. Our starting
point is an EFT action describing the  two BSs (labelled by $A=1,2$)
as worldlines $\mathrm{\mathbf{z}}_A(\tau_A)$ endowed with time-dependent complex
sources $s_A(\tau_A)$,
\begin{align}
  &S_{\rm EFT} = \int \frac{\sqrt{-g}}{2}\left\{ \frac{R(g)}{8\pi G}
        -\left[ g^{\mu\nu}\nabla_{\mu}{\bar{\varphi}}
        \nabla_{\nu}\varphi
        +\mu^2 \bar{\varphi} \varphi\right]\right\}\du^4x
        \nonumber\\
        &~-\sum_A \int \Big\{m_A - 2\pi\Big[\varphi(\mathrm{\mathbf{z}}_A) \bar{s}_A(\tau_A)
        +\bar{\varphi}(\mathrm{\mathbf{z}}_A) {s}_A(\tau_A)\Big]\Big\} \du \tau_A\,
        \nonumber\\
        & + S_{\rm GF} (g).
  \label{eq:actionEFT}
\end{align}
Here, $S_{\rm GF}(g)$ fixes the harmonic gauge and $\tau_A$ is proper time. 
We keep only the quadratic term $\mu^2 \bar{\varphi} \varphi$ in
the scalar potential because the higher-order terms $\sim
\varphi^4+\varphi^6$ lead to exponentially faster decaying interactions.
The complex sources describing BSs are harmonically varying, i.e.
\begin{align}
  s_A(\tau_A) =  c_A e^{\iu \epsilon_A \,\omega_A\tau_A + \iu \Phi_A} ~,
  \label{eq:Scharge}
\end{align}
with constant (real) ``scalar charges" $c_A$. The nonlocal character of $s_A$ \Big(via $\tau_A = \int^{z_A} \sqrt{-g_{\mu\nu}(z'_A)dz^{'\mu}_Adz^{'\nu}_A}$\Big) can be described by using a Lagrange multiplier. The harmonically varying
nature of the scalar sources $s_A(\tau_A)$ is a crucial new feature
compared to existing EFT descriptions of point-like objects interacting
in scalar-tensor theories of gravity. 

\begin{figure}[t]
 \subfloat[]{
\begin{tikzpicture}[scale=0.8]
  \draw[line width=0.3mm] (-3.2,0.3) -- (-2.2,0.3);
  \draw[line width=0.15mm, scalar] (-2.7,0.3) -- (-2.7,-1.0);
  \draw[line width=0.3mm] (-3.2,-1.0) -- (-2.2,-1.0);
\end{tikzpicture}
\label{fig:1a}
}%
\hfill
\subfloat[]{
\begin{tikzpicture}[scale=0.8]
  \draw[line width=0.3mm] (-1.9,0.3) -- (-0.7,0.3);
  \draw[boson] (-1.7,0.3) -- (-1.3,-0.35);
  \draw[scalar] (-1.3,-0.35) -- (-0.9,0.3);
  \draw[scalar] (-1.3,-0.35) -- (-1.3,-1.0);
  \draw[line width=0.3mm] (-1.9,-1.0) -- (-0.7,-1.0);
\end{tikzpicture}
\label{fig:1b}
}%
\hfill
\subfloat[]{
\begin{tikzpicture}[scale=0.8]
  \draw[line width=0.3mm] (-1.9,0.3) -- (-0.7,0.3);
  \draw[scalar] (-1.7,0.3) -- (-1.3,-0.35);
  \draw[scalar] (-1.3,-0.35) -- (-0.9,0.3);
  \draw[boson] (-1.3,-0.35) -- (-1.3,-1.0);
  \draw[line width=0.3mm] (-1.9,-1.0) -- (-0.7,-1.0);
\end{tikzpicture}
\label{fig:1c}
}%
\hfill
\subfloat[]{
\begin{tikzpicture}[scale=0.8]
  \draw[line width=0.3mm] (-1.9,0.3) -- (-0.7,0.3);
  \draw[boson] (-1.3,0.3) -- (-1.7,-1.0);
  \draw[scalar] (-1.3,0.3) -- (-0.9,-1.0);
  \draw[line width=0.3mm] (-1.9,-1.0) -- (-0.7,-1.0);
\end{tikzpicture}
\label{fig:1e}
}%
\vspace{-0.5em}
\caption{Feynman diagrams needed for the computation of the
scalar field effective action up to $\mathcal{O}\left( G c_A c_B\right)$.
The dashed (wavy)  line represents the scalar (graviton)  propagator.}
\label{fig:feynman-all}
\end{figure}
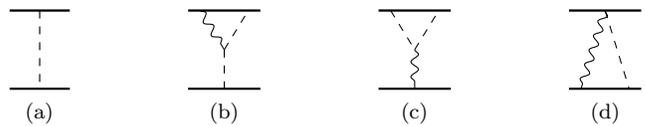

The perturbative analysis from this EFT leads to Feynman-type
diagrams computing the impulse $ \Delta p_A^{\nu} \defeq  p^\nu_{A
\, \rm out} -  p^\nu_{A \,\rm in}$ of each body under the exchange
of the scalar field together with gravity. The diagram Fig.~\ref{fig:1a}, describing scalar exchange between
$s_1$ and $s_2$, yields a LO contribution to the
impulse {of body one} given by
\begin{align} \label{impulse}
  \Delta p^{\nu \, \rm scalar}_{1 \rm\, LO} &= -\frac{8\pi \tilde{\mu} c_1 c_2 }{\sqrt{\gamma^2-1}} {K_1\big(\tilde{\mu} |b|\big)} \frac{b^{\nu}}{|b|} \cos(\Phi_{21} )
  \nonumber\\
  &-
  \frac{8\pi c_1 c_2}{\sqrt{\gamma^2-1}} (\omega_1 \gamma-\omega_2) \check{u}_2^{\mu}\sin(\Phi_{21} ){K_0(\tilde{\mu}}|b|)~.
\end{align}
Here, $\check{u}_2^{\mu}= (u_2^{\mu}-u_1^{\mu}\gamma)/(\gamma^2-1)$, $b^{\nu}= b_1^{\nu}-b_2^{\nu}$ is the impact parameter,
$K_1, K_0$ are modified Bessel's functions, and $\widetilde{\mu}$  is
as above.  {The first (transverse) term on the right-hand side of \eqref{impulse} corresponds (using $j= p_{\infty} |b|/(GMh)$) to the
scattering angle} $\chi_{\rm LO}^{\rm scalar}$ ,
\begin{align}
   \frac{\chi_{\rm LO}^{\rm scalar}(j)}{h}= \frac{8\pi  M\widetilde{\mu}}{m_1 m_2}
   \frac{ c_1 c_2}{\gamma^2-1} K_1\left(\tilde{\mu}
   |b|\right) \cos(\Phi_{21}) ~.
   \label{eq:Scalarchi}
\end{align}
{The second (longitudinal) term on the right-hand side of \eqref{impulse} predicts that the scattering is not conservative when $\Phi_{2} \neq \Phi_1$. We have checked this prediction against NR data.
This non conservative nature of the scattering can be analytically described by completing the conservative potential $w^{\rm scalar}$ by a dissipative force ${\cal F}^{\rm scalar}_{\mu \, A}$ acting on each body. We checked on NR data, however, that the nonconservative nature of the scalar interaction introduces a  negligible difference (for the cases we consider) between the scattering angle of particle 1, $\chi_1^{\rm NR}$, and that of particle 2, $\chi_2^{\rm NR}$, so that we can define $\chi^{\rm NR} \defeq \frac12 (\chi_1^{\rm NR}+\chi_2^{\rm NR})$. We henceforth discuss the construction of a scalar contribution, $w^{\rm scalar}(\bar r)$, in a radial potential \eqref{eq:wdecompose} leading to $\chi^{\rm NR}$.
}
Expanding Eq.~\eqref{eq:chirelation} to linear order in $w$ implies
that the scalar contribution to $\chi$ is related to the 
scalar contribution to $w$ via the following linear transformation
\begin{align}
\label{chivswscalar}
  \chi^{\rm scalar}(j) = -\frac{\partial}{\partial j}
  \int_{\bar{r}_{\rm min}}^{\infty} \du \bar{r}\frac{ w^{\rm scalar}
  (\bar{r})}{\sqrt{p_{\infty}^2-j^2/\bar{r}^2}}~.
\end{align}
One readily checks that inserting in Eq.~\eqref{chivswscalar} the Yukawa-type potential $w_{\rm LO}^{\rm
scalar}$, \eqref{eq:wscalar}, gives rise to $\chi_{\rm LO}^{\rm
scalar}$, \eqref{eq:Scalarchi}. 

To go beyond LO in the scattering angle, one needs to include both higher-order diagrams and iterated effects of the perturbed worldlines. The most relevant higher-order diagrams for our AR/NR comparison are \ref{fig:1b}-\ref{fig:1e} which describe a $\mathcal{O}(G)$ gravitational dressing of Fig.~\ref{fig:1a}.
{Restricting ourselves to the equal-mass case, implying $c_1 = c_2$, $m_1 = m_2$ and $\omega_1 = \omega_2 = \omega$, we infer from matching the EFT results to
the exact isolated, gravitationally dressed BS solution
(see Eq.~\eqref{eq:bckgA} below) that the }  leading fractional modifications of Eq.~\eqref{eq:Scalarchi}
for large impact parameter are logarithmic and of order $G m_A ((2\omega^2-\mu^2)/\widetilde{\mu}) \log(2e^{\gamma_{\textrm{E}}}\widetilde{\mu}b)$. In the absence of exact results for the $O(G c_A c_B)$ contributions, we shall assume that the corresponding higher-order contribution to $w$ can be approximated by the factorised form
\begin{align}
  w^{\rm scalar}_{\rm HO} = w_{\rm LO}^{\rm scalar}
  w_{\operatorname{g-dressing}}^{\rm scalar}~,
  \label{eq:wHOinit}
\end{align}
with the following exponentiated gravitational dressing factor:
\begin{align}
  w_{\operatorname{g-dressing}}^{\rm scalar} = e^{G M ((2\omega^2-\mu^2)/\widetilde{\mu})  \log(2e^{\gamma_{\rm E}}
  GM\widetilde{\mu}\bar{r})}\,.
  \label{eq:wHO}
\end{align}
%

We use the integral \eqref{eq:chirelation}, with
$w=w^{\rm BH}  + w^{\rm tid}  + w^{\rm scalar}_{\rm HO}$, to
define the analytic prediction for the scattering angle.

{\textit{\textbf{Completing the AR description}}---}
The AR-predicted angle \eqref{eq:chitotal} depends on the knowledge
of the following quantities: $E$ (or equivalently $\gamma=2(E/M)^2-1$),
$J$, $w_5^{\rm BH}$, $w_6^{\rm tid}$, $w_7^{\rm tid}$, the
scalar-charge parameters $c_1 (=c_2)$, and the phase difference $\Phi_{21}$. We discuss their derivation in turn.

Residual constraint violations inherent to our initial data
construction limit the accuracy of computing the initial values of
$E$ and $J$ from their standard integral expressions.  As discussed
in the appendix \ref{append1} the approximate
analytical expression \eqref{EN} with $c_{\rm V}=1.37$, which yields,
\begin{equation} \label{Ein}
   \frac{ E}{M} = 1.02394~~~\text{with}~~~ \gamma_{\rm in} =1.09690\,,
\end{equation}
 defines a sufficiently
accurate value of the initial energy; we adopt it in this work.
In the absence of an analytic expression for $J$ analogous to
Eq.~\eqref{EN} (i.e.~incorporating $\sim \frac{G m}{2 X}
=\mathcal{O}(10^{-2})$ fractional corrections), we allow for a correction $J= (1 +
\epsilon_J) J_{\rm NR}$ to the ADM-like initial angular
momenta listed in Table \ref{tab:initData}, and determine best-fit
values for $\epsilon_J$ in the following way. Since the scalar
interactions average out for BS$\overline{{\rm BS}}$ configurations
their evolution most closely resembles BH scattering. We consequently
use the subset of five BS$\overline{{\rm BS}}$ simulations with
$b/(GM)\le13.0$
together with analytic information about the scattering of BHs to
fit $\epsilon_J = 0.006 \pm 0.001$ and, hence, use for {\it all}
simulations
\begin{align}
\label{deltaJ}
  J =(1+0.006) \, J_{\rm NR} \,.
\end{align}

Interpolating between the results of Refs.~\cite{Damour:2014afa,
Swain:2024ngs}, we have determined the effective 5PM contribution
to the BH potential $w^{\rm BH}({\bar r}, \gamma)$  corresponding
to our initial energy \eqref{Ein},
\begin{align}
  w_5^{\rm BH} = -1.03 \pm 0.03~.
\end{align}

Next, we use the results of Ref.~\cite{Sennett:2017etc} to estimate
the dimensionless even-parity quadrupolar tidal parameter $\Lambda_A
\defeq \lambda^{\rm tidal}_A/(G^4 m_A^5)$, where $\lambda^{\rm tidal}_A$ denotes the tidal deformability of our solitonic BSs, and
obtain $\Lambda_A  \approx 7$ for each model.  In the absence of
corresponding results for the odd-parity quadrupolar and the
even-parity octupolar tidal parameters, we insert $\Lambda_A  \approx
7$ in Refs.~\cite{Bini:2020flp,Cheung:2020sdj, Kalin:2020lmz} to
determine $\chi_6^{\rm tid}$ and $\chi_7^{\rm tid}$ and then
use Eq.~(\ref{chitidal}) to obtain the corresponding tidal coefficients
$w_6^{\rm tid}$, and $w_7^{\rm tid}$ entering the EOB $w$-potential.

The scalar charges $c_A$ are determined by matching the (gravitationally
dressed) EFT-predicted analytic expression of the asymptotic scalar
profile generated by an isolated BS in harmonic
coordinates, namely (recalling ${\bar \mu_A}
\defeq \sqrt{\mu^2 - \omega_A^2}$),
\begin{equation}
  A(r) \approx \frac{c_A}{r}   \exp \left[- {\bar \mu_A} r + G m_A
  \frac{ 2 \omega_A^2 - \mu^2}{\bar \mu_A} \log(2e^{\gamma_{\rm E}} {\bar \mu_A} r)
  \right].
  \label{eq:bckgA}
\end{equation}
to the scalar profile numerically computed for isolated BSs.  For
our models with $G\mu m_A = 0.713$ and $\omega_A=0.439 \mu$ we obtain
%
  $\sqrt{G} \mu  c_A = {1.608}\,.$
%

%
%


{\textit{\textbf{Comparison between AR and NR results}}---}
We now compare our AR predictions $\chi_{\rm AR}$, defined by
Eq.\eqref{eq:chirelation}, with our full set of 32 NR scattering data $\chi_{\rm AR}$ (see Table \ref{tab:initData})
composed of four types of BS binaries, each simulated
for eight values of $J$ (or, equivalently, $b$).

\begin{figure}[t]
  \centering
  \includegraphics[width=\linewidth]{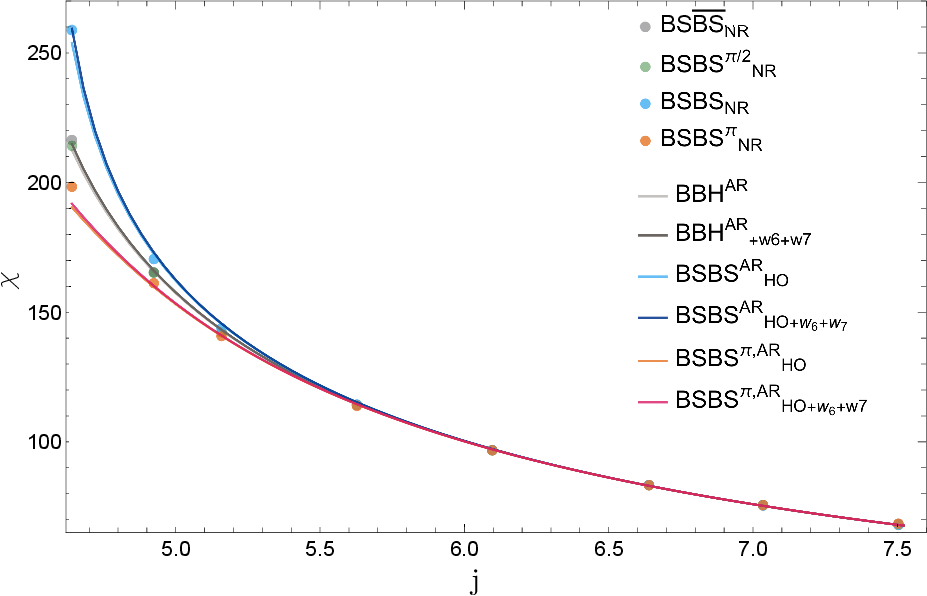}
  \caption{
  Scattering angle comparison between the NR data (filled circles) and the
  EOB-resummed, PM based analytical prediction (AR; solid lines) for equal-mass,
  non-spinning boson star configurations for various rescaled initial
  angular momentum ($j =J/(G m_1 m_2)$).
  }
  \label{fig:compareNRvsAR}
\end{figure}

{Fig.~\ref{fig:compareNRvsAR} presents a global view of the comparison between our AR predictions (solid
lines)  and the corresponding  NR data (circles).}  For all four sequences, the
agreement is visually excellent for most impact parameters, and exhibits a small worsening
only for the smallest impact parameters (leading to very large
scattering angles, $\chi > 200^{\circ}$). {Our AR predictions are presented both in their full version (using $w=w^{\rm BH}  + w^{\rm tid}  + w^{\rm scalar}_{\rm HO}$)
and a version using $w=w^{\rm BH}  + w^{\rm scalar}_{\rm HO}$.}

\begin{figure}[t]
  \centering
  \includegraphics[width=\linewidth]{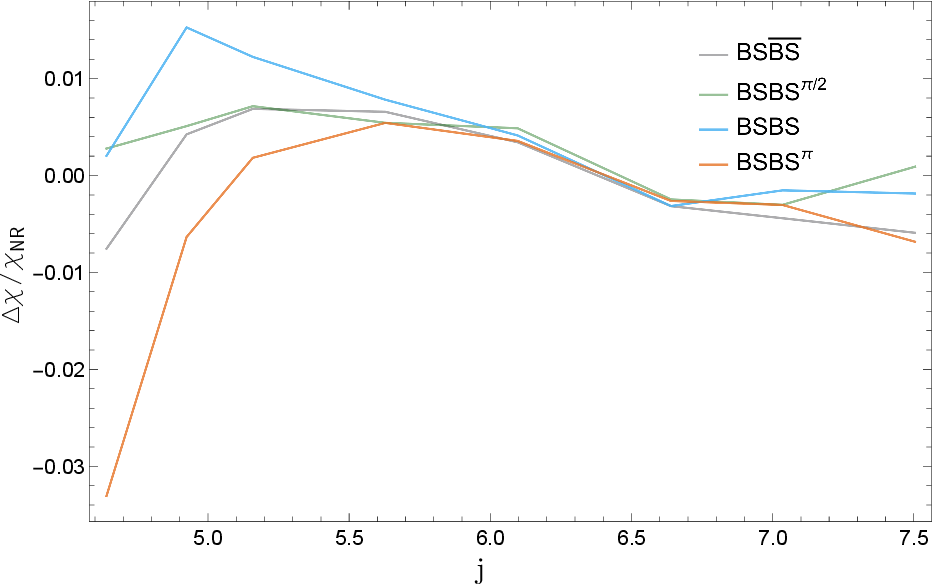}
  \caption{
  Fractional differences $\Delta\chi/\chi_{\rm NR} = (\chi_{\rm AR}-\chi_{\rm NR})/\chi_{\rm NR}$ of analytical results (including both the
  tidal and scalar interactions) with respect to the numerical
  results.}
  \label{fig:compareNRvsAR1}
\end{figure}

We quantify the {(percent level)} AR-NR agreement
by displaying in Fig.~\ref{fig:compareNRvsAR1} the fractional differences
$(\chi_{\rm AR} - \chi_{\rm NR})/ \chi_{\rm NR}$. 
Further details about the NR/AR comparison are given in
Table~\ref{tab:ARdata}, which explores the effect of various versions
of the radial potential such as including only LO scalar interactions
and/or various tidal terms.  There, we define the quantity $\delta_{\chi} \defeq (\chi_{\rm AR} - \chi_{\rm NR})/\Delta
\chi_{\rm NR}$ that measures the deviation of AR
from NR values in units of the NR error estimate $\Delta \chi_{\rm NR}$. Note that $\delta_{\chi}$ is of order unity for our most accurate AR models.


The use of an analytic description based on $w^{\rm EOB}$-resummed
angles is crucial to reach this level of agreement; PM-expanded
angles would exhibit up to an order of magnitude larger differences
from NR data for nearly all impact parameters.
The best agreement is obtained for the BS$\rm{\overline{BS}}$
sequence (black lines in Fig.~\ref{fig:compareNRvsAR}) which, as
expected, is very close to BH-binary scattering; that agreement was obtained by fitting only one parameter, namely $\epsilon_J$. As indicated, tidal effects are small even for small
impact parameters, $b= 9.90 M, 10.5 M$.  The agreement for the
BSBS$^{\frac{\pi}{2}}$ sequence (green symbols in Fig.~\ref{fig:compareNRvsAR})
is nearly as good, and also close to the BBH case. Its slightly
larger disagreement for small $b$ (leftmost data points in
Fig.~\ref{fig:compareNRvsAR1}) likely arises because the vanishing
of the $w$ potential for BSBS$^{\frac{\pi}{2}}$ is only valid at LO,
i.e.~only for the diagram Fig.~\ref{fig:1a}.  Excellent agreement
with NR data is also obtained for the other two configurations, (i)
BSBS (blue), and (ii) BSBS$^{\pi}$ (red).  These cases
also clearly demonstrate the attractive (repulsive) character of the
scalar-mediated potential $w^{\rm scalar}$, when $\Phi_{21}=0$
($\Phi_{21}=\pi$). In the attractive BSBS case, the two stars
get very close to each other, {$\bar{r}_{\rm min} \approx 6.8$}, and $w_{\rm HO}^{\rm scalar}$, Eq.~\eqref{eq:wHOinit}, becomes quite significant;
cf.~Table~\ref{tab:ARdata}.

\begingroup\squeezetable
\begin{table*}[t]
\begin{center}
\caption
{Summary of the AR scattering angle for
various EOB predictions with impact parameter $b$. Here,
$\delta_{\chi} = (\chi_{\rm AR} - \chi_{\rm NR})/\Delta
\chi_{\rm NR}$ measures the deviation of AR
from NR values in units of the NR error estimate.
}
\begin{tabular}{ | c | c | c | c | c | c | c | c |c | c | c | c | c | c | c | c| c| c| c |c | c |}
\hline
$\frac{b}{GM}$ & $\chi^{\rm BBH}_{\rm AR}\,[{}^{\circ}]$ & $\delta_{\chi}^{\rm BS\overline{BS}}$ & $\delta_{\chi}^{\frac{\pi}{2}}$\!\! & \multicolumn{4}{c|}{$\chi^{\rm scalar}_{{\rm LO}, \rm AR}\,[{}^{\circ}]$} & \multicolumn{4}{c|}{$\chi^{\rm scalar}_{{\rm HO},\rm AR}\,[{}^{\circ}]$} & \multicolumn{4}{c|}{$\chi^{\rm scalar}_{{\rm HO}, w_6, \textrm{AR}}\,[{}^{\circ}]$} & \multicolumn{4}{c|}{$\chi^{\rm scalar}_{{\rm HO}, w_6+w_7, \textrm{AR}}\,[{}^{\circ}]$} \\
\hline
& & & & BSBS & $\delta_{\chi}$  & BSBS$^{\pi}$ & $\delta_{\chi}^{\pi}$ & BSBS & $\delta_{\chi}$ & BSBS$^{\pi}$ & $\delta_{\chi}^{\pi}$ & BSBS & $\delta_{\chi}$ & BSBS$^{\pi}$ & $\delta_{\chi}^{\pi}$ & BSBS & $\delta_{\chi}$ & BSBS$^{\pi}$ & $\delta_{\chi}^{\pi}$\\
\hline
\hline
9.9  &  212.44 & -2.81 & -1.24 &  -     & -      & 110.41 & -69.84  & 253.68  & -5.38 & 190.44 & -6.33 & 256.24 & -2.69 & 191.12 & -5.79 & 259.32 & 0.56  & 191.89 & -5.20\\
10.5 &  165.62 & 0.25  & 0.42  & -      & -      & 115.45 & -61.04  & 172.51  & 2.25  & 159.78 & -1.94 & 172.83 & 2.61  & 160.01 & -1.63 & 173.11 & 2.93  & 160.21 & -1.36 \\
11.0 &  143.10 & 1.30  & 1.36  & 371.98 & 367.86 & 115.60 & -43.50  & 145.47  & 2.50  & 140.90 & 0.12  & 145.58 & 2.69  & 141.00 & 0.30  & 145.67 &  2.84 & 141.09 & 0.45\\
12.0 &  114.90 & 0.12  & 1.08  & 124.70 & 16.06  & 107.50 & -9.93   & 115.27  & 1.32  & 114.53 & 0.90  & 115.31 & 1.37  & 114.56 & 0.93  & 115.32 & 1.40  & 114.58 & 0.96 \\
13.0 &  97.21  & 0.38  & 0.76  & 99.16  & 2.70   & 95.39  & -1.73   & 97.28   & 0.46  & 97.14  & 0.40  & 97.29  & 0.47  & 97.15  & 0.42  & 97.30  & 0.48  & 97.15  & 0.42 \\
14.1 &  83.12  & -0.23 & -0.19 & 83.46  & 0.07   & 82.76  & -0.50   & 83.12   & -0.23 & 83.10  & -0.20 & 83.12  & -0.23 & 83.10  & -0.20 & 83.13  & -0.23 & 83.10  & -0.20 \\
15.0 &  75.39  & -0.24 & -0.18 & 75.50  & -0.01  & 75.28  & -0.27   & 75.39   & -0.10 & 75.39  & -0.18 & 75.40  & -0.10 & 75.39  & -0.18 & 75.40  & -0.10 & 75.39  & -0.18 \\
16.0 &  68.03  & -0.28 & -0.05 & 68.06  & -0.08  & 68.01  & -0.34   & 68.03   & -0.10 & 68.03  & -0.32 & 68.03  & -0.10 & 68.03  & -0.32 & 68.04  & -0.10 & 68.03  & -0.32 \\
\hline
\end{tabular}

\label{tab:ARdata}
\end{center}
\end{table*}
\endgroup\squeezetable

Our results also display the sub-dominant effect due to electric-type,
quadrupolar tidal interactions. The inclusion, order by order, of
the tidal potentials $w_6^{\rm tid}$,  $w_7^{\rm tid}$, while nearly
negligible for medium and large $b$, improves the agreement with
NR data for all configurations with small impact parameters
Appendix \ref{sec:tidalEffects}.

{\textit{\textbf{Conclusions}}---}
We have performed the first numerical computation of the scattering
angle $\chi$ of nonspinning, equal-mass BS binary encounters
with scattering angles between 68.2$\degree$ and
259$\degree$.
Our key results include: (i) the derivation of the
analytical LO scalar-exchange contribution to the scattering based
on the EFT action \eqref{eq:actionEFT} involving harmonically-varying
scalar sources, (ii) inclusion of the gravitational-dressing effects
to complete the LO scalar interactions with HO effects, and (iii)
determination of the effective scalar charges $c_A$ by matching to
NR computations of isolated BSs to EFT predictions.

The comparison of the numerical/analytical results for BSs shows
(i) the dominant effect due to scalar interactions, and (ii) the subdominant but real
effect due to tidal interactions when the two stars get close. This
indicates that the nature of the constituent matter of \textit{exotic}
compact objects can contribute more significantly to the dynamics
than the tidal interactions.

Our simulations also
revealed that for small impact parameters both BSs collapse into BHs
after their close encounter and before separating to infinity. This
phenomenon was not
observed in quasi-circular BS binaries with the same compactness,
highlighting the potentially important role of time-dependent scalar
perturbations in destabilizing BSs and triggering collapse. This
opens a novel direction for probing smoking gun-signatures of exotic
compact objects.

%
%

\begin{acknowledgments}
T.~D.  and T.~J. thank Romain Gervalle for confirming the scalar
charge of boson-stars. T. J. is supported by a LabEx Junior Research
Chair Fellowship at ENS, Paris. T.~J. thank the hospitality and the
stimulating environment of the Institut des Hautes Etudes Scientifiques
(IHES).  The present research was partially supported by the 2021
Balzan Prize for Gravitation: Physical and Astrophysical Aspects,
awarded to T. Damour.  This work has been supported by STFC Research
Grant No. ST/V005669/1.  We acknowledge support by the NSF Grant
Nos.~PHY-090003,~PHY-1626190 and PHY-2110594, DiRAC projects ACTP284
and ACTP238, STFC capital Grants
Nos.~ST/P002307/1,~ST/R002452/1,~ST/I006285/1 and ST/V005618/1,
STFC operations Grant No.~ST/R00689X/1. Computations were done on
the CSD3 and Fawcett (Cambridge), Cosma (Durham), Stampede2 (TACC)
and Expanse (SDSC) clusters.
\end{acknowledgments}

\appendix

\twocolumngrid

\section{Calculation of Initial Energy}
\label{append1}
\renewcommand{\theequation}{A.\arabic{equation}}
\setcounter{equation}{0}
There are two features of our numerical setup that {\it a priori}
affect the accuracy of our analytical results: (i)  the non-Cartesian
value of the asymptotic spatial metric $\gamma^{\infty}_{ij} \neq
\delta_{ij}$; and (ii) constraint violations in the initial data.

The feature (i) has a nearly negligible effect on our results for
the following reason. The asymptotic spatial metric of the improved
superposition of two boosted single-BS spacetimes with BS
distance $r_{AB}$ is $\gamma^{\infty}_{ij} = 2
\delta_{ij}-\gamma^{B}_{ij}(x_A^i) $; cf.~Eqs.~(45)
in \cite{Helfer:2021brt}. To a good approximation this
is $\gamma^{\infty}_{ij} \approx \left(1- 2 G m/r_{AB}  \right)
\delta_{ij}$, which differs from a Cartesian metric by a factor
$a^2 = 1- 2 G m/r_{AB}=1- G M/r_{AB}$.  Such a factor in the
asymptotic metric can be reabsorbed in a common rescaling of the
units of length, time and mass by a factor $a$ (this leaves $G$ and
$c$ fixed). Such a rescaling changes, for instance, the numerical
value of the ADM mass-energy (with respect to using the usual
Cartesian-based integral expression of the ADM energy) as
$E_{\rm ADM}^{\gamma^{\infty}_{ij}}= E_{\rm ADM}^{\delta_{ij}}/a$, which,
for our initial data ($a^2 \approx 1- 1\%$), means an increase of
$E_{\rm ADM}$ by about $+ 0.5 \%$.  However, this common rescaling of
units of length, time and mass leaves invariant all dimensionless
quantities (such as $E/M$, $J/(G M^2)$ and $\chi$) and
therefore can be ignored when comparing the function $\chi^{\rm
AR}(E/M, J/(G M^2))$ to $\chi^{\rm NR}$; cf.~Sec. 3.1 of
\cite{DD86} for a similar discussion of the irrelevance of the
Doppler effect in binary pulsar gravitational tests.

The second feature (constraint violations) impacts  the ADM-like
evaluation of $E$ from the initial time slice ($t=0$)
without affecting the BS masses. It thereby affects the numerical
value of the dimensionless ratio $E/M$, which plays a
crucial role in our  AR/NR comparison. When ignoring the rescaling
factor linked to $\gamma^{\infty}_{ij} \neq \delta_{ij}$, i.e.~when
computing the ADM-like initial energy (at $t=0$) by using Eq.~(7.15)
of Ref.~\cite{Gourgoulhon:2007ue} (which assumes $\gamma^{\infty}_{ij}
= \delta_{ij}$) one obtains (exactly) the sum of the ADM integrals
for two boosted Schwarzschild metrics,
\begin{equation} \label{eq:MadmtZeroAnal}
  E^{\rm NR}_{\rm ADM} (t =0) =  E_{\rm kin}(v)=1.027324 M\,.
\end{equation}
Here the numerical value is obtained by inserting $v=0.2291$ into
the special-relativistic expression for the kinetic energy $E_{\rm kin}$ of two moving bodies,
\begin{equation} \label{eq:Ekin}
  E_{\rm kin} \defeq \frac{m_A}{\sqrt{1-v_A^2}}+  \frac{m_B}{\sqrt{1-v_B^2}}
  = \frac{2m}{\sqrt{1-v^2}}\,.
\end{equation}
Extrapolating our numerical evaluation of the ADM integral to
infinite radius yields $ E^{\rm NR}_{\rm ADM} (t =0) = (1.027095
\pm 0.000361) M$ compatible with Eq.~\eqref{eq:MadmtZeroAnal}.

The value \eqref{eq:MadmtZeroAnal}, however, represents only the
kinetic energy of two moving bodies and does not account for the
system’s gravitational binding energy. A more accurate value of the
total incoming energy of the system differs from the kinetic-energy
\eqref{eq:Ekin}, by an additional term approximately equal to the
Newtonian potential energy,
\begin{equation} \label{eq:VN}
  V_N=  - \frac{G m_A m_B}{r_{AB}}= - \frac{G m^2}{2 X} \approx -0.002476 M~.
\end{equation}
We indeed obtain such a correction by taking into account the effect
of the initial violations in the Hamiltonian constraint which we
write as
\begin{equation} \label{eq:H0}
  \mathcal{H}_0 \defeq \mathcal{R} + K^2 - K_{mn}K^{mn} 
  -16\pi \rho \neq 0\,.
\end{equation}
Here $\mathcal{R}$ is the Ricci scalar of the spatial metric
$\gamma_{ij}$, $K_{ij}$ is the extrinsic curvature with trace $K$,
and $\rho$ denotes the scalar energy density. Defining the unphysical
energy-density contribution
\begin{equation}
  \rho_u \defeq \frac{\mathcal{H}_0}{16 \pi}\,,
\end{equation}
Eq.~\eqref{eq:H0} can be rewritten as a normal-looking Hamiltonian constraint,
\begin{equation}
  \mathcal{H} \defeq \mathcal{R} + K^2 - K_{mn}K^{mn}  
  -16\pi (\rho + \rho_u)=0\,.
\end{equation}
The latter equation shows that our initial data correspond to a
total matter-energy density equal to the sum $\rho_{\rm tot}=\rho
+ \rho_u$, rather than featuring only the physical energy density
$\rho$.  We must therefore {\it subtract} the effect of the unphysical
energy density $\rho_u$ from our ADM-computed initial mass-energy
\eqref{eq:MadmtZeroAnal}.  We can estimate this unphysical contribution
to the ADM mass-energy by integrating $\rho_u$ over the initial
hypersurface $\Sigma$,
\begin{equation} \label{eq:Mu}
  E_{\rm u} = \int_{\Sigma} \rho_u \sqrt{\gamma}\,\du x^3\,,~~~
  \gamma \defeq \det \gamma_{ij}\,.
\end{equation}
For the BS binary configurations of Table \ref{tab:ARdata}, a
numerical estimation of the integral \eqref{eq:Mu} yields\footnote{We
obtain the same result by analyzing otherwise identical initial
data for binary separations $X/M=60$, $80$ and $100$, indicating
the relation's more generic validity.}
\begin{equation} \label{eq:Munum}
  E_{\rm u} = (1.5 \pm 0.2) V_N\,,
\end{equation}
and a numerical estimate of the total energy given by
\begin{equation} \label{eq:Ekinu}
    E^{\rm NR} = E^{\rm NR}_{\rm ADM}(t=0) - E_{\rm u} = E_{\rm kin} - 1.5 V_N\,.
\end{equation}
This result is qualitatively confirmed by computing the ADM integral
as a function of time and bearing in mind the constraint damping
of the CCZ4 formulation; our code involves (\`a la Ref.~\cite{Alic:2011gg})
a constraint damping parameter $\kappa_1=0.1/(GM)$. The time evolution
$E_{\rm ADM}^{\rm NR}(t)$ indeed exhibits an early decrease
commensurate with Eq.~(\ref{eq:Ekinu}), albeit with some numerical
noise arising from the discretization error incurred by time evolving
the data.

Analytic arguments independently validate our expression for the
initial energy as follows. We start by assuming that the total
incoming energy of the binary can be written as
\begin{align}
  \label{eq:Einc}
  E(c_V) = E_{\rm kin} + c_V V_{\rm N}~,
\end{align}
and study the effect of changing the value of the coefficient $c_V$
in front of  $V_N$.  From post-Newtonian theory, we know that $c_V
= 1 + \mathcal{O}(v^2) + \mathcal{O}(Gm/X)$. For our initial
configuration we can neglect $\mathcal{O}(Gm/X)$ compared to
$\mathcal{O}(v^2)$. When considering the BH scattering
data of Ref.~\cite{Damour:2014afa} (and evaluating  $ E_{\rm kin}$
as a function of the incoming momenta), one finds that their $ E_{\rm
in}^{\rm NR}$ is
recovered for $c_V \approx 1+ 7 v^2$.  [The value of $c_V$ depends
both on the choice of coordinates, and on the use of a Lagrangian
versus an Hamiltonian formalism. For instance, the usual
first-order post-Newtonian conserved energy in the Lagrangian approach
would feature $c_V = 1-7v^2$.] Formally applying the expression $c_V \approx
1+ 7 v^2$ to our case (i.e. $v=0.2291$) yields $c_V \approx 1.37$
in good agreement with Eq.~(\ref{eq:Ekinu}).

\begin{figure}[t]
  \centering
  \includegraphics[width=\linewidth]{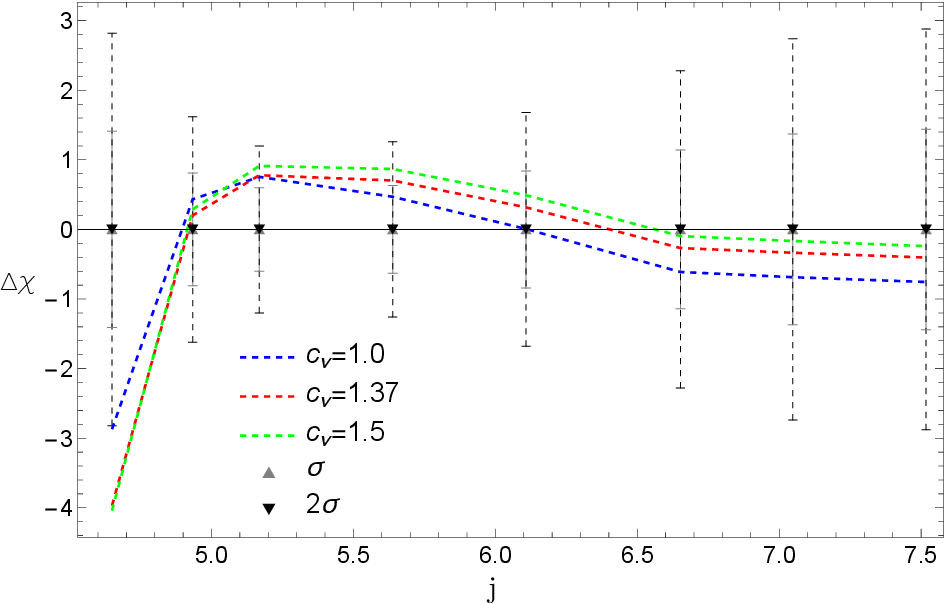}
  \caption{
  Difference $\Delta \chi=\chi^{\rm AR} - \chi^{\rm NR}$ between
  the analytic and NR values of the scattering angle for
  BS$\rm{\overline{BS}}$ binaries with initial
  energy given by Eq.~(\ref{eq:Einc}) with $c_V=1$, $c_V=1.37$ and $c_V=1.5$.
  The error bars ($\sigma$) correspond to the NR results presented
  in Table \ref{tab:initData}.}
  \label{fig:upEn1}
\end{figure}

We now show that defining $E$ by Eq. \eqref{eq:Einc}, with
different coefficients $c_V$, leads to essentially equivalent results for the
purpose of our AR/NR comparison of scattering angles.  In the main
text we compare the scattering data  of $\rm{BS\overline{BS}}$
binaries to $\chi^{\rm AR}(E/M, (1+\epsilon_J) J_{\rm
NR}/(G M^2))$ using  $E= E({c_V=1.37})$, and fitting
for $\epsilon_J$. The results of the same analysis using $E= E(c_V=1.0)$ (for which we found $\epsilon^{c_V=1.0}_J
= 0.008$) and $E= E(c_V=1.5)$ (with $\epsilon^{c_V=1.5}_J
= 0.005$) are shown in Fig.~\ref{fig:upEn1} and Table
\ref{tab:upEn}. By comparing with Fig.~\ref{fig:compareNRvsAR}
(where $c_V=1.37$), we see that choosing either $c_V=1$ or  $c_V=1.37$ or $c_V=1.5$
has only a minor impact on the scattering angle except for the
smallest impact parameter $b=9.9 G M$.  Indeed, the  differences
$\chi^{\rm AR}- \chi^{\rm NR}$ measured in units of the NR error
bars, namely $\delta_{\chi}\defeq (\chi^{\rm AR}- \chi^{\rm
NR})/\Delta \chi^{\rm NR}$, remain within approximately two standard
deviations for all the cases $c_V=1$, $c_V=1.37$ and $c_V=1.5$ (indicated as
labels).

\begin{table}
\caption
{Summary of the fractional deviations ($\delta_\chi$) of analytic results from numerical results
(in units of the NR error bars) for BS${\rm \overline{BS}}$ binaries with the three energy values $E(c_V=1)$, denoted by $\delta_{\chi,1}$, for $E(c_V=1.37$) denoted by $\delta_{\chi,1.37}$, and for $E(c_V=1.5$) denoted by $\delta_{\chi,1.5}$ .
}
\begin{center}
\footnotesize{
\begin{tabular}{c  c  c   c  c }
\hline
\hline
$b/(GM)$ & $\delta_{\chi,1}$ & $\delta_{\chi,1.37}$ &  $\delta_{\chi,1.5}$ & \\
\hline
\hline
9.9  & -2.03 & -2.81 & -2.86\\
10.5 &  0.53 &  0.25 & 0.36\\
11.0 &  1.26 &  1.30 & 1.52\\
12.0 &  0.74 &  1.12 & 1.38\\
13.0 &  0.01 &  0.38 & 0.58\\
14.1 & -0.54 & -0.23 & -0.08\\
15.0 & -0.50 & -0.24 & 0.12\\
16.0 & -0.52 & -0.28 & -0.17\\
\hline
\hline
\end{tabular}
}
\label{tab:upEn}
\end{center}
\end{table}

\section{Impact of tidal interactions}
\label{sec:tidalEffects}
\renewcommand{\theequation}{B.\arabic{equation}}
\setcounter{equation}{0}

Here, we illustrate the improvement obtained by including tidal
effects for small impact parameter. In order to optimally isolate
the tidal effects, we consider the BS$\rm{\overline{BS}}$ configurations with minimal scalar interaction.
Table \ref{tab:tidesBBH} lists the analytically
predicted scattering angles including the tidal interactions for
these configurations; these scattering angles complement the
corresponding values reported in the 2nd and 3rd columns of Table \ref{tab:ARdata}
computed {\it without} tidal effects.  While the deviation
$\delta_{\chi}^{\rm{BS\overline{BS}}}$ remains largely
unaffected by the inclusion of tidal effects for $b\ge 10.5\,GM$,
we observe a significant reduction in the deviation from NR results
from $3\sigma$ to $1\sigma$ for the smallest impact parameter
$b=9.9\,GM$.  This improvement confirms that tidal effects become
significant when the two stars get close to each other. A similar
comparison for the BSBS$^{\frac{\pi}{2}}$ sequence (cf.~the 4th column of
Table \ref{tab:ARdata}) exhibits less improvement after addition
of tidal interactions compared to BS$\rm{\overline{BS}}$ sequence,
probably because the scalar interactions vanish only at leading
order for this binary type.

\begin{table}[t]
\caption
{Summary of the analytically predicted scattering angle along with
the deviation from NR results including tidal interactions for
BS$\mathrm{\overline{BS}}$ and for BSBS$^{\frac{\pi}{2}}$
(denoted by $\frac{\pi}{2}$) configurations. For all configurations, the total energy is $E_{\rm
in}/M =  1.02394$.
}
\footnotesize{
\begin{tabular}{c| c c c| c c c}
\hline
$b/(GM)$  &  $\chi^{\textrm{BBH}+w6}_{\rm AR}$ & $\delta_{\chi}^{{\rm BS}\overline{\rm BS}}$ & $\delta_{\chi}^{\frac{\pi}{2}}$ & $\chi^{\textrm{BBH}+w6+w7}_{\rm AR}$ & $\delta_{\chi}^{{\rm BS}\overline{\rm BS}}$ & $\delta_{\chi}^{\frac{\pi}{2}}$  \\
\hline
\hline
9.9   & 213.55 & -2.03 & -0.45  & 214.78 & -1.16 &  0.43 \\
10.5  & 165.89 &  0.58 & 0.75  & 166.13 &  0.87 &  1.04 \\
11.0  & 143.21 &  1.49 & 1.55  & 143.30 &  1.64 &  1.70 \\
12.0  & 114.93 &  1.16 & 1.14  & 114.95 &  1.20 &  1.18 \\
13.0  & 97.22 &  0.39 & 0.78  & 97.22  &  0.40 &  0.79 \\
14.1  & 83.12  & -0.23 & -0.19 & 83.11  & -0.23 & -0.19 \\
15.0  & 75.39  & -0.24 & -0.18 & 75.39  & -0.24 & -0.17 \\
16.0  & 68.03  & -0.28 & -0.06 & 68.03  & -0.28 & -0.06 \\
\hline
\hline
\end{tabular}
}
\label{tab:tidesBBH}
\end{table}

\section{Varying the initial separation}
\renewcommand{\theequation}{C.\arabic{equation}}
\setcounter{equation}{0}
In order to further test the performance of our analytic model,
we compare here the the AR and NR scattering angles for 5 additional
simulations of BS$\rm{\overline{BS}}$ binaries with fixed impact
parameter $b=10.5\,GM$ but different initial separations. The initial
energy is computed from Eq.~\eqref{eq:Ekinu} and our results are
presented in Table ~\ref{tab:NewPoints}. The excellent agreement
of $\sim1 \sigma$ or better demonstrates the accuracy of our model independent
of the initial separation (provided it is large enough to properly
define the incoming angle).
\vspace{6pt}
\begin{table}[H]
\caption
{Initial separation $X$, NR scattering angle, analytic scattering
for $E$ as given by Eq.~(\ref{eq:Einc}) with $c_V=1.37$ and deviations $\delta_{\chi}$ for the BS-$\rm{\overline{BS}}$
binary sequences. For all configurations, the impact parameter is
$b=10.5\,GM$.
}
\begin{center}
\footnotesize{
\begin{tabular}{c c c c}
\hline
\hline
$X/M$   &  $\chi^{\rm BS\overline{BS}}_{\rm NR}$ & $\chi^{\rm{BBH}}_{\rm AR}$ & $\delta_{\chi}$ \\
\hline
\hline
50.48  & 165.42(81) & 165.62 & 0.25 \\
60.30  & 166.08(71) & 166.48 & 0.56 \\
70.12  & 166.39(58) & 167.12 & 1.26 \\
79.94  & 166.76(56) & 167.62 & 1.53 \\
89.75  & 167.09(55) & 168.01 & 1.68 \\
100.27 & 167.39(55) & 168.36 & 1.76 \\
\hline
\hline
\end{tabular}
}
\label{tab:NewPoints}
\end{center}
\end{table}
%

\bibliography{prlref}
\onecolumngrid
\renewcommand{\theequation}{SM.\arabic{equation}}
\setcounter{equation}{0}

\end{document}